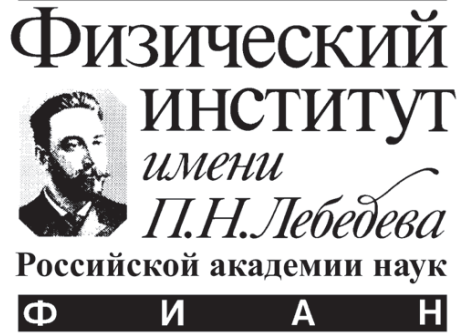

S. FRANCHINO, M. NEGODAEV, A. BOLSHAKOV,
E. ASHKINAZI, Y. KALKAN, A. POPOVICH,
M. KOMLENOK, V. SOSNOVTSEV, V. RALCHENKO

# GAS ELECTRON MULTIPLIER BASED ON LASER-PERFORATED CVD DIAMOND FILM: FIRST TESTS

# Gas electron multiplier based on laser-perforated CVD diamond film: First tests

S. Franchino[a*], M. Negodaev[b**], A. Bolshakov[c,d], E. Ashkinazi[c,d], Y. Kalkan[e], A. Popovich[c,f], M. Komlenok[c], V. Sosnovtsev[d], V. Ralchenko[c,d],

[a] *Kirchhoff Institut fur Physik, Ruprecht-Karls-Universitat Heidelberg, Germany*
[b] *Lebedev Physical Institute RAS, Moscow, Russia*
[c] *General Physical Institute RAS, Moscow, Russia*
[d] *National Research Nuclear University MEPhI, Moscow, Russia*
[e] *Muş Alparslan University, Muş, Turkey*
[f] *Institute of Radio Engineering and Electronics RAS, Fryazino, Russia*

## ABSTRACT

Gas electron multiplier (GEM) is widely used in modern gas detectors of ionizing radiation in experiments on high-energy physics at accelerators and in other fields of science. Typically the GEM devices are based on a dielectric foil with holes and electrodes on both sides. GEMs made by radiation-hard dielectrics or wide band-gap semiconductors are desirable for some applications. The results of the first tests of the gas electron multiplier made of radiation-hard materials, such as polycrystalline CVD diamond with a thickness of 100 microns is described. Here we report on fabrication of GEM based on free-standing polycrystalline CVD diamond film and its first test.

---

[*] Corresponding author. E-mail: Silvia.Franchino@cern.ch
[**] Corresponding author. E-mail: negodaev@lebedev.ru

The use of microelectronic technology in the production of gas position-sensitive detectors has led to the development of a family of novel devices called micro-pattern gas detectors (MPGD), including a wide topologies of geometries. Among them one could mention, in chronological order, Microstrip Gas Chamber (MSGC) [1], Microgap Chamber [2], Microdot Chamber [3], Micro-Mesh Gas structure (Micromegas) [4], Gas Electron Multipliers (GEM) [5], Micro-Groove Detector [6], etc.

The main feature of those detectors is the spatial separation between a conversion region (called also drift region), where primary electrons are created by electromagnetic interactions between the incoming ionizing particles and the gas molecules, and amplification regions, where gas avalanche processes happen. The typical distance between the electrodes in the amplifying cell is 10-100 µm. This allows to the MPGD to operate at high rates, up to $10^6$ particles/mm$^2$s [7], (thanks to the very small distances between anode and cathode, hence the fast removal of positive ions by nearby cathode strips, reducing the space-charge build up, typical of wire chambers) and exhibit excellent spatial (~ 30 µm, determined by the diffusion of electrons in the gas) and multitrack resolution (~ 500 µm). Recent aging studies revealed that they might be even less vulnerable to the radiation damage effects, compared to standard silicon microstrip detectors, if reasonable precautions are taken on the components quality [8].

In MPGDs the electrodes of the amplifying cells are typically produced by a lithography process and are located on a dielectric substrate. The presence of the insulator near the electrodes leads to a gradual change of the electric field due to the insulator charging, and to dependence of the gas gain on the rate.

The first MPGD detector ever studied, the MSGC, has shown to be prone to electrical breakdown at high rates and in the presence of high ionizing particles. Although these devices usually withstand the breakdown without damage, they demand protection of the sensitive front-end electronics. In addition, the recovery time which is required after the device gets such breakdowns can severely affect the efficiency of the charge registration.

In 1997 F. Sauli (CERN) solved the problems of breakdowns in the MSGC proposing to insert an additional element into the gas volume, which he called gas electron multiplier (GEM) [5].

GEM foils are produced is a thin composite with two metal layers separated by a thin insulator, and etched with a regular matrix of conical holes with a high density (50-100 holes per mm$^2$). When applying the potential difference to the electrodes of the GEM, each holes works as an independent proportional counter; electrons formed in the drift region on one side of the structure move into the holes and are multiplied in the channels under the high electric field.

MSGC detectors with GEMs as preamplifier element were used in the inner tracker of the HERA-B experiment at DESY [9, 10].

As demonstrated later, the GEMs can operate not only combined with the MSGCs, but also can be used themselves for the detection of ionizing radiation, if it is coupled to a read-out electrodes board that collect the amplified charge.

Nowadays most of GEM detectors are designed to be operated in multiple cascaded stages, typically three; in this way the process of amplification is shared into these stages, and the electric field strength on the dielectric surface is reduced, and reduced is also the discharge probability, keeping the necessary gas gain (usually $10^3$-$10^4$) for effective detection.

In the 1990-ies LPI in collaboration with the SINP MSU worked on the creation and investigations of radiation-hard detectors with high spatial resolution for high-energy physics experiments. In particular, the position-sensitive gas detectors were considered. Together with the ORION company and the enterprises of Zelenograd microstrip gas chambers with different semiconductive coatings and gas electron multipliers were produced [11-14]. The manufacturing technology of GEM production was different from the technology at CERN. In 1999 the first GEM with a gain of 5000 was produced, which allowed the authors to propose to use GEM together with readout strips as the gas position-sensitive detector [9].

Detectors with three stages of multiplication by GEM (size 31x31cm$^2$) successfully operate for the first time at high rates in the COMPASS experiment at CERN [15].
In 2008, the collaboration RD51 was organized at CERN, which aims in developing technologies for production of microstructure detectors and for their use in experiments in high energy physics and in further applications. GEM detectors are deeply studied inside this collaboration and are considered as very promising. Nowadays most of the high energy particle physics experiments are using, or are planning to use in the incoming upgrades, GEMs as part of their detector equipment, especially in the forward regions, where the particle rates and radiations are higher than the one that classical wire chambers can withstand.
As an example one could mention GEM detectors in most of the LHC experiments at CERN: TOTEM tracker [16], LHCb [17] and CMS (upgrade) [18] for the innermost trigger layer of the muon spectrometer and ALICE [19] that will soon replace the wire chambers in the TPC with 4 stages GEM detectors (for an efficient ion backflow suppression).
GEM detectors are not only used in particle physics experiment, but many R&D studies are progressing in many other fields, like plasma diagnostics and medicine [20,21] or neutron detection [22], muon tomography in homeland security [23] and many others.

Up to date the classical GEM foils that are used in high energy physics experiments are produced, mostly at CERN, using a chemically etchable polyimide film (50 μm thickness), copper-plated on both sides (5 μm). However, for some specific applications, there are some disadvantages in using this material, like the degassing and the changing of its resistance with high radiation ion beam [24, 25]. Moreover, since the assembly and the stretching of such flexible foils is performed manually, working performance of these detectors can have a strong variation due to the influence of the human factor.

In recent years GEM R&D studies on the production of GEM detector prototypes with different dielectric materials are ongoing, with the advantage to have flat and self-sustaining surfaces without stretching, cylindrical holes less affected by charging up effects and the absence of outgassing materials. Ceramic or glass material have been successfully implied for GEM prototypes [26].

Radiation-hard materials like diamond, with stable resistivity are also being studied for stable operation of the GEM detectors in very high radiation environments, like, for example, the very forward regions of high energy physics detectors or as a beam or luminosity monitor.

In this paper we describe the first development and tests of a GEM detector prototype with, as a dielectric material, the polycrystalline film of chemical vapor deposition (CVD) diamond. The CVD diamond films is chemically stable, extremely mechanically and radiation hard and with very high thermal conductivity [27, 28]. While mechanical treatment of polycrystalline diamond is very difficult, its laser cutting and drilling is well developed [29], and was used in the present work for holes perforation.

The resistivity of CVD diamond films can be tuned by the film growth parameters and may vary within wide range even without any doping. A sort of defective CVD diamond ("black" diamond) with a reduced resistivity of $10^9$-:-$10^{10}$ Ω•cm seems to be appropriate for the GEM fabrication.

The first test sample of the diamond GEM was prepared at the General Physics Institute RAS, in collaboration with LPI RAS, using a "black diamond" film ≈100 μm thick. The mother wafer of polycrystalline black diamond wafer of ≈100 μm thickness has been grown on in $CH_4(10\%)/O_2(1\%)/H_2$ mixture in a microwave plasma CVD system ARDIS-100 (5 kW microwave power, 2.45 GHz frequency) on a Si substrate of 57 mm in diameter [30]. After chemical etching of the Si substrate the resulting free-standing diamond wafer a 15 × 15 $mm^2$ segment has been cut from it by a Nd:YAG laser. At the next step the array of 50 μm diameter holes the inter distance of 120 μm was produced by drilling with a copper-vapor laser on an area of 10x10 $mm^2$ (Fig. 1).

The electrodes of the test GEM structure on both sides were then produced by graphitization of the diamond surface layer by irradiation with ArF excimer laser (193 nm wavelength, 20 ns pulse width). The laser beam scanned over the film surface to ablate the top layer of diamond and form a thin conductive graphitic layer, which will serve as electrode. To remove the graphite from the hole's walls the sample was treated in *aqua regia*, $HNO_3$ (65%) - HCl (35%), that increased the resistance by four orders of magnitude from $8*10^4$ kΩ to $5*10^8$ kΩ, corresponding to final resistivity of $5*10^{13}$ Ω·cm.

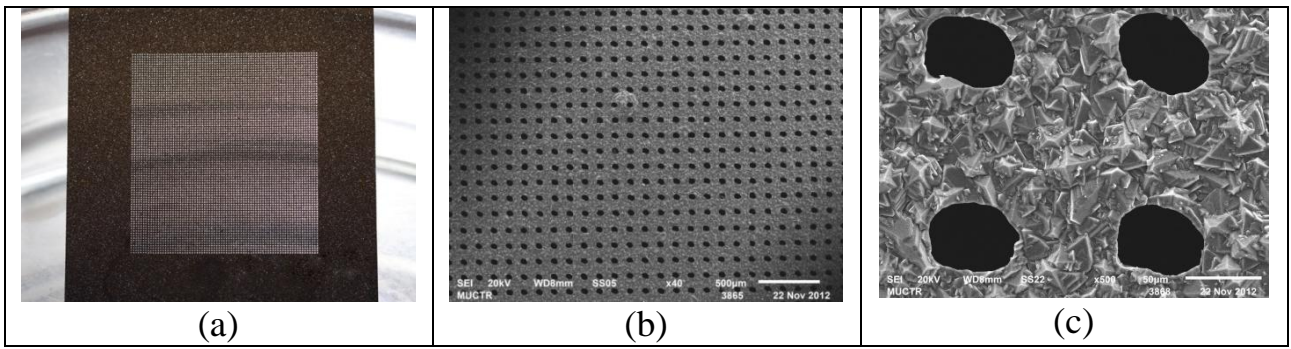

(a)　　　　　　　　　　(b)　　　　　　　　　　(c)

Fig. 1. The photograph of the diamond plate (15x15 mm$^2$) with an array of holes in central part of the plate (a) and SEM images of a selected area with the holes at different magnifications (b,c).

The graphitic layer formed in similar manner by laser treatment revealed good performance as Ohmic contacts in diamond particle detectors in our previous work [31].

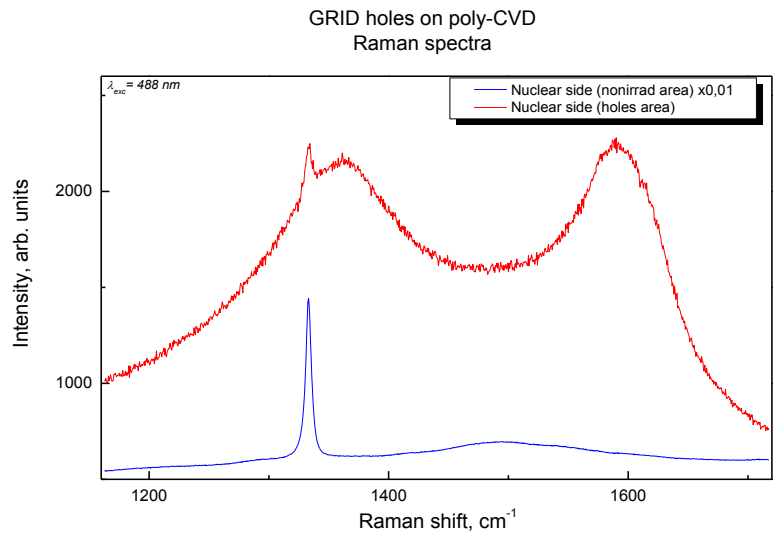

Fig. 2. Raman spectra taken on virgin diamond film (bottom) and on area graphitized by Ar excimer laser irradiation (top). The appearance of broad D-band (1360 cm$^{-1}$) and G-band (1600 cm$^{-1}$) indicates diamond-to graphite transformation on the irradiated surface.

The occurrence of graphitization was confirmed with micro-Raman spectroscopy using LABRAM HR 800 (Horiba Jobin Yvon) instrument at 488 nm excitation wavelength by focusing Ar$^+$ laser beam in a spot of ~ 1 μm in diameter at the film surface. While for virgin film the only feature in the Raman spectra were the narrow peak at 1332.5 cm$^{-1}$ belonging to diamond phase and weak wide band at 1500 cm$^{-1}$ from amorphous carbon inclusions, strong broad D-band (1360 cm$^{-1}$) and G-band (1600 cm$^{-1}$) appeared for laser-treated surface, indicating diamond-to graphite transformation on the irradiated area (Fig. 2).

In October 2015, preliminary performance tests have started at the Gas Detector Development laboratory of CERN.

The GEM sample was assembled together with a drift electrode (D) made with a stainless steel mesh, and an anode (A) made with copper tape.

The detector prototype was assembled such that the drift gap was 6 mm thick and the induction gap (distance between the GEM and the anode) was 3 mm. In order to apply voltage to the two sides of the diamond GEM, two electrodes have been created during the assembling phase.

The top side of the GEM was mounted on a wide copper foil with a quadratic hole that corresponded with the active area of the detector (~ 1 cm$^2$), and then connected to the High Voltage line (GEM TOP). Electrical contact between the copper foil and the surface of the graphitized part of GEM was provided by conductive glue.

On the bottom side, the graphitized surface of the GEM was connected by conductive glue, to a copper wire. This electrode was called GEM BOT.

A scheme of a chamber with the GEM is shown in Fig. 3 and in Fig. 4, the photographs of assembly steps of the chamber for testing are shown.

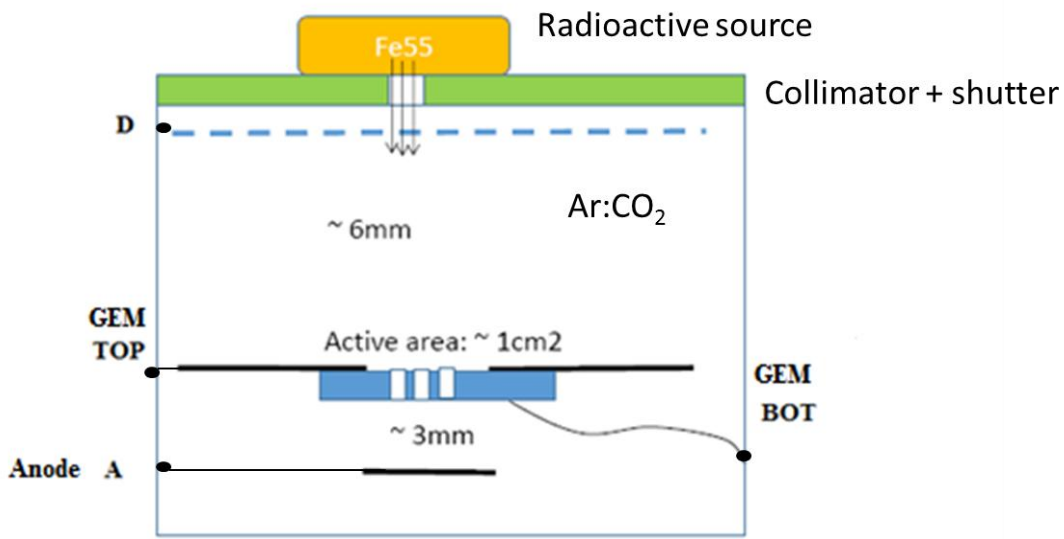

Fig. 3. A transversal scheme of the setup for diamond GEM tests.

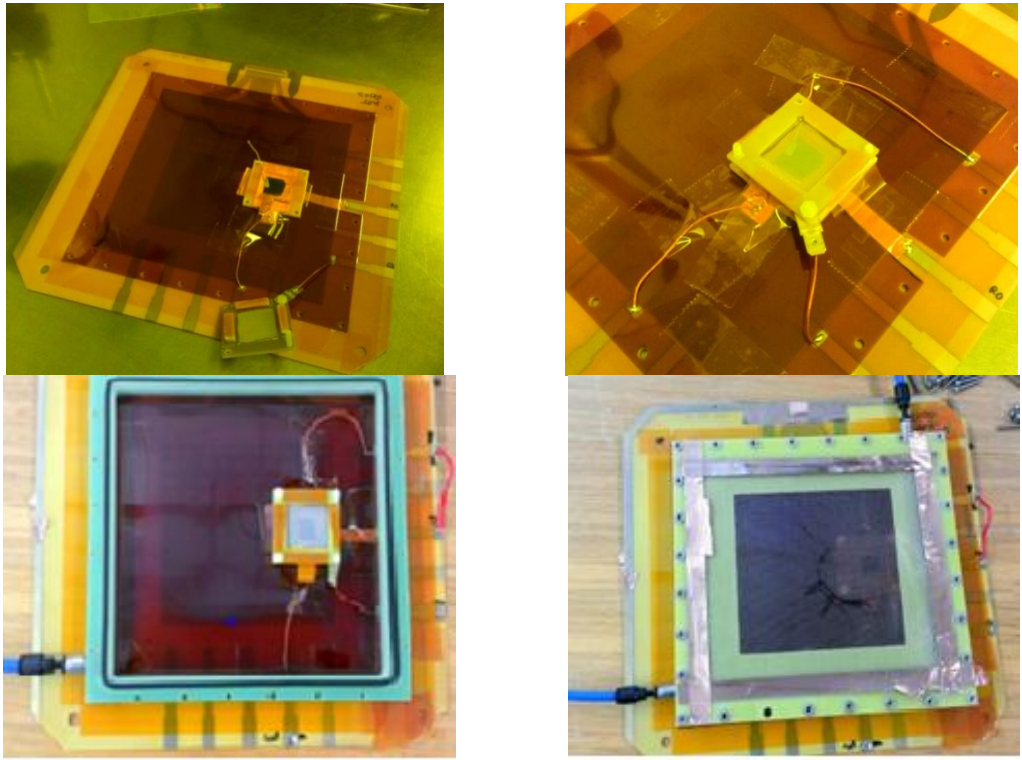

Fig. 4. Photographs of assembly steps of the chamber for testing CVD diamond GEM

The stack was placed inside a gas volume with connectors to ensure the flow of the gas mixture used for the tests and the powering of the four electrodes. As gas mixture we have used Ar:CO2 93:7 with flow at about 3 l/h.

First, in order to be sure of the electrical connection between the electrodes, and in order to measure the Ohmic behavior of this material and its leakage current, we measured the current across the two electrodes when a difference of potential was applied between the GEM TOP and GEM BOT electrodes.

The I-V curve is shown in Fig. 5. From the fit of the experimental points we obtain a resistance of $10^{14}$ Ohm, close to the value measured immediately after perforation and acid treatment. The current was measured with a pico ammeter Keithley 6487 (100fA current resolution), and we took care of don't exceed the 400V as potential difference across the GEM in order to be sure to don't start the amplification process inside the gas.

The detector was subsequently irradiated with a radioactive source of $Fe^{55}$ (1,11 GBq) and we measured the current collected at the anode in dependence on the voltage at the electrodes of the GEM.

A collimator with a hole diameter of ~1cm$^2$ was used in order to ensure the ionization of the gas only on top of the active area of the GEM detector. The radioactive source was fixed on top of this collimator in order to irradiate the same zone for all the measurement.

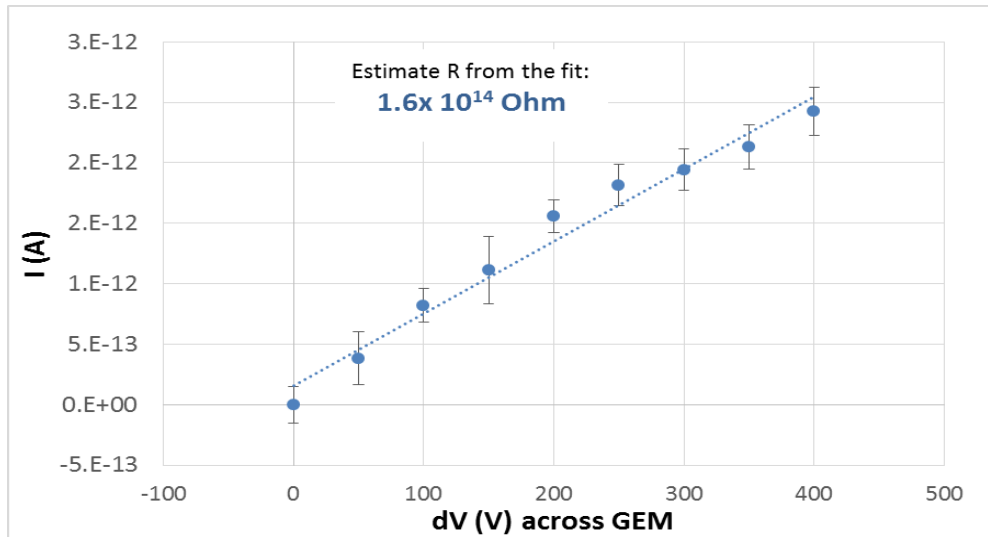

Fig. 5. I-V curve of the diamond GEM and the estimation of its resistance through the fit of the experimental points.

In order to simplify the system, we decided to eliminate the induction field; the detector was equipped again with a pico-ammeter Keithley 6487 that was connected directly to the electrode GEM BOT, that was grounded through a protection resistor of 10 MOhm.

In order to subtract the baseline fluctuations from the pico-ammeter, we define as a current collected on the anode (Ia) the difference between the current collected with the radioactive source open and the one with the source closed with a shutter installed on top of the collimator.

In Fig. 6 with the green circles, one can see how the anode current (Ia) varies as a function of the potential difference across the GEM (dV) with the setup that has been described above (setup a).

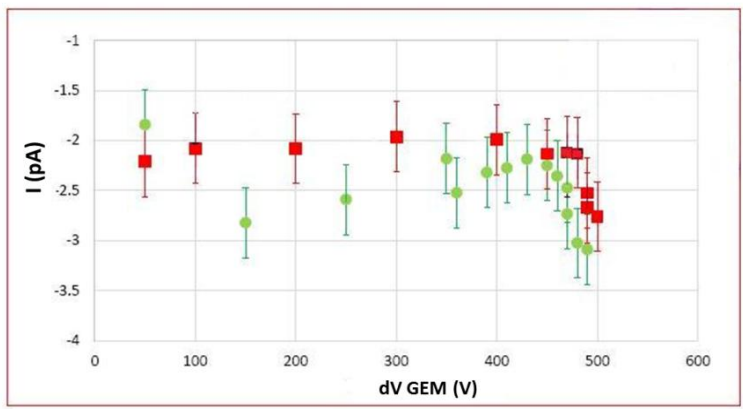 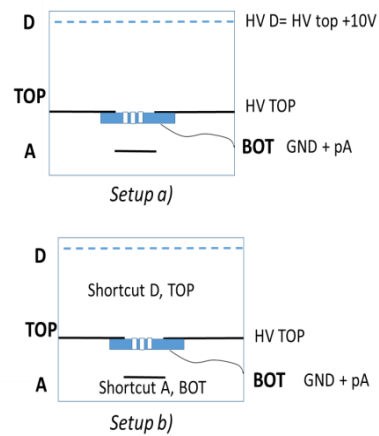

● − a)  ■ − b)

Fig. 6. The dependence of the current measured on GEM for different schemes of connection.

In this configuration, in order to maximize the collection of primary electrons inside the GEM holes and avoid electric field lines to drop on top of the non-active area, the drift field was very low (as it is later shown in Fig. 7). For each point of the scan, the drift voltage was set at 10V above the voltage of GEM TOP.

From the results shown in Fig 6 one can see that at voltages across the GEM higher than 450 V, the collected current slightly increases as a result of the GEM gas amplification. At 500 V between GEM electrodes, the amplification factor is about 1.5. With this measurements we couldn't exceed voltages across the GEM above 500 V because instabilities of current were observed, possibly due to discharges in the GEM. Such discharges could occur due to various reasons, such as, for example, the inhomogeneity of the surface resistivity on the GEM electrode, the presence of not graphitized areas with a large resistance on the electrode surface, the presence of graphitized areas in the holes or dust, or because some defects in the laser-drilling process of this first prototype (as it can be seen from Fig. 1c).

Due to the not optimal geometry and collection efficiency of our setup (too small active area with respect to the drift electrode and drift gap) we believed that very little of the primary charge created in the conversion region was focused inside the GEM holes and then amplified. Most of that could be captured by the relatively big electrode GEM TOP. In order to demonstrate that, we decided to slightly change the setup (setup b in Fig 6) in which the powering schema of the device was modified in order to remove the effects of the primary charge produced in the drift region. In this case the drift field was kept to zero for all the scan, thanks to a shortcut to the top electrode; the bottom electrode was shorted to the collector electrode, where the pico-ammeter was connected. In this case the collected charge at the GEM BOT electrode is only due to primary ionization created very close to each GEM holes. The red squares in Fig 6 are the results of this measurement and indeed the behavior is very similar to the one of the previous scan.

In order to remove the hypothesis that the seen current was coming from the bulk material of the GEM under direct irradiation, we have repeated both measurements changing the gas of the detector and fluxing pure $CO_2$. In both cases the measured current was compatible with zero for every voltages applied across the GEM.

In Fig. 7 it is shown the result of the collected current on the BOT GEM as a function of the drift field, when a voltage across the GEM of 450V is applied.

One can see that in order to eliminate the collection of the primary charge and then amplification inside the GEM holes, a reverse drift field of >500V/cm need to be applied. On the other hand, due to the geometry of the hole pattern and of the not optimal electrode connections, one can see that as soon as the drift field exceeds ~500V/cm the measured current decreases. This is probably due to the drop of electric field lines on top of the top GEM electrode. In this case, primary electrons created in the drift region, are not focused inside the GEM holes, but are lost on the GEM electrode. With the geometry of this prototype the optimal drift field is indeed very small, compatible to zero value, as it was used in the previous measurement.

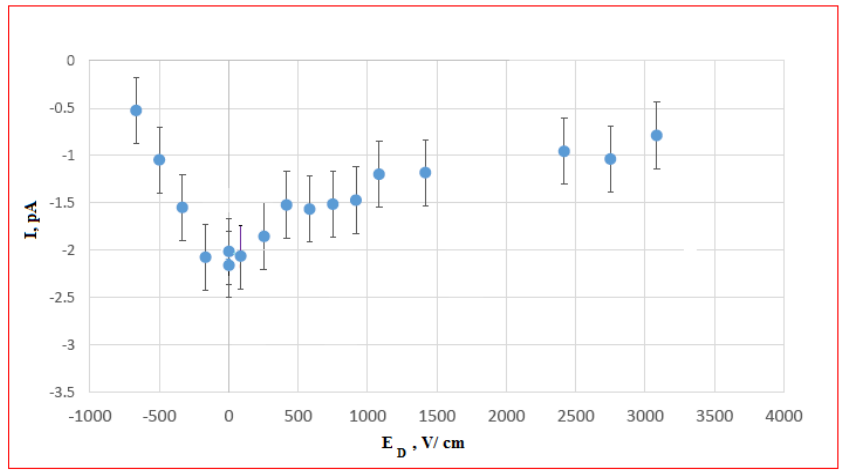

Fig. 7. Dependence of the current in the circuit electrode BOT GEM on the electric field in the drift gap at a voltage between the GEM electrodes of 450 V.

In Fig. 8, the monitoring of the measured current as seen with the pico-ammeter is shown, with the radioactive source being closed (upper points) and open (lower points) with the shutter. Each point of the previous scans is taken doing the mean value over 100 measured points (both for the closed and open source); the error bars are given by the standard deviation.

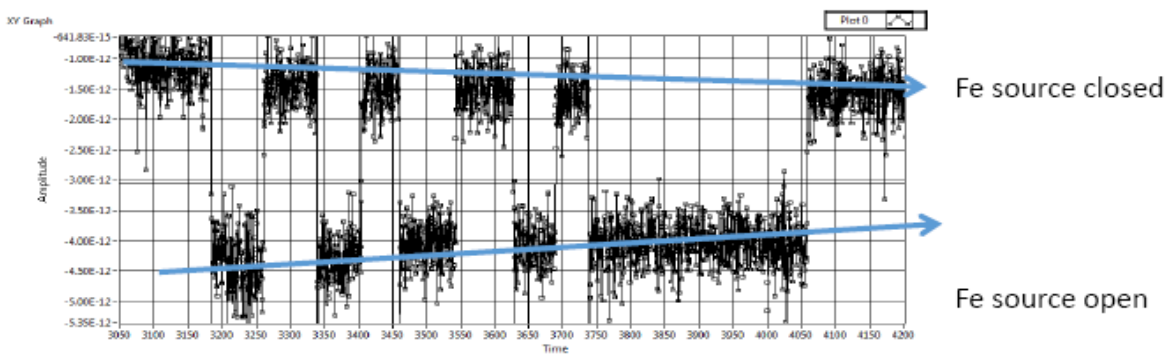

Fig. 8. Online current measurements at the GEM electrode and at the anode with closed and opened source $Fe^{55}$.

A current drift was observed at repeated measurements with the same electrode voltages, which might be associated with a polarization defects in the "black" CVD diamond and subsequent possible charging up of the surface (if, for example, the graphitization process in creating the electrodes was not perfect).

After analyzing the results of the first tests of the CVD diamond GEM working performance it was decided to produce an additional CVD diamond GEM samples, with the electrodes produced by the graphitization of the polycrystalline CVD-diamond plate surface, and with electrodes made by metallization of the diamond-plate surface. Two samples of diamond grid with conical holes (with input and output diameters of 60 μm and 90 μm, respectively, and a step of 120 μm) are already prepared (Fig. 9).

Further improvement of technology of manufacturing CVD diamond plates with a large number of micron-sized holes will allow to build, on the basis of GEM structures, micro-patterned gas detectors for detection of ionizing particles with stable operation in high-radiation environment.

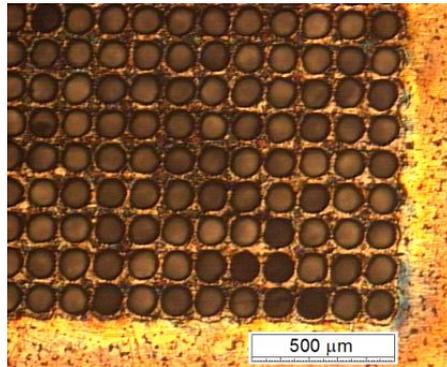

Fig. 9. Optical microscope image of a part of the diamond grid with 100 μm diameter holes.

**Acknowledgements**

The authors are thankful to A. Khomich for measurement of Raman spectra, V. Paramonov for the work on laser drilling, A. Romaniouk and V. Tikhomirov for interest and support of GEM production, and to the Gas Detector Development group at CERN (in particular E. Oliveri and L. Ropelewski) for having hosted us during our measurements and for their precious suggestions.